\documentclass[12pt,njp]{iopart}
\usepackage[pdftex]{graphicx}
\usepackage{bm}
\usepackage{epsfig}
\usepackage{dsfont}
\usepackage{bbm}
\usepackage{amsfonts}
\usepackage{amssymb}
\usepackage{amscd}
\usepackage{amstext}
\usepackage{float}
\usepackage{color}
\usepackage{ulem}
\usepackage{enumerate}
\usepackage{iopams}
\usepackage[export]{adjustbox}

\newcommand{\ket}[1]{|\,{#1}\,\rangle}

\usepackage[noadjust,compress]{cite}
\newcommand{\kvec}{\mathbf{k}}
\newcommand{\xvec}{\mathbf{r}}

\newcommand{\expec}[1]{\langle #1 \rangle}

\def\beq{\begin{equation}}
\def\eeq{\end{equation}}

\def\CR{\nonumber\\[0.15cm]}

\newcommand{\rref}[1]{Ref.~\cite{#1}}
\newcommand{\frefp}[2]{Fig.~\ref{#1}~(#2)}

\newcommand{\esref}[2]{Eqs.~(\ref{#1}) and (\ref{#2})}

\newcommand{\cref}[1]{chapter~\ref{#1}}
\newcommand{\Cref}[1]{Chapter~\ref{#1}}

\newcommand{\bref}[1]{(\ref{#1})}

\usepackage{ulem}  
\newcommand{\fdd}[0]{\Phi_{\rm dd}}
\newcommand{\bds}{\boldsymbol}

\begin{document}

%
%
%
\title{Anisotropic Inflation in Dipolar Bose-Einstein Condensates}
\author{A.~Rana$^1$, A.~Pendse$^2$, S.~W\"uster$^{1}$ and S.~Panda$^{1}$}
\address{$^1$Department of Physics, Indian Institute of Science Education and Research, Bhopal, Madhya Pradesh 462 066, India}
\address{$^2$Max Planck Institute for the Physics of Complex Systems, N\"othnitzer Str. 38, 01187 Dresden, Germany}
\ead{arunrana@iiserb.ac.in, sebastian@iiserb.ac.in}
\begin{abstract}
Early during the era of cosmic inflation, rotational invariance may have been broken, only later emerging as a feature of low-energy physics. This motivates 
ongoing searches for residual signatures of anisotropic space-time, for example in the power spectrum of the cosmic microwave background.
We propose that dipolar Bose-Einstein condensates (BECs) furnish a laboratory quantum simulation platform for the anisotropy evolution of fluctuation spectra during inflation, 
exploiting the fact that the speed of dipolar condensate sound waves depends on direction.
We construct the anisotropic analogue space-time metric governing sound, by linking the time-varying strength of dipolar and contact interactions in the BEC to the scale factors in different coordinate directions. Based on these, we calculate the dynamics of phonon power spectra during an inflation that renders the initially anisotropic universe isotropic.
We find that the expansion speed provides an experimental handle to control and study the degree of final residual anisotropy.
Gravity analogues using dipolar condensates can thus provide tuneable experiments for a field of cosmology that was until now confined to a single experiment, our universe.
\end{abstract}

\maketitle

\section{Introduction} 
The cosmological principle, the assumption that our universe is isotropic and homogeneous on the largest length scales, is strongly supported by the isotropic thermal microwave radiation field known as Cosmic microwave background (CMB).  But as we zoom in closer, we find several unexpected features \cite{WMAP:2006bqn,Copi:2006tu,Bennett:1996ce,WMAP:2006jqi,Akrami:2014eta,Planck:2018jri,Planck:2019evm} in the CMB, such as the alignment of lowest multipoles \cite{WMAP:2003ivt,Copi:2013cya}, a hemispherical power asymmetry \cite{Eriksen:2004iu,Hoftuft_2009}, a preference for odd parity modes \cite{PhysRevD.72.101302,PhysRevD.82.063002,Kim_2010} and a large cold spot in the southern hemisphere \cite{Vielva:2003et,Cruz:2004ce,Larson_2004}. There are several mechanisms to explain their origin \cite{Schwarz:2015cma}, one of which involves primordial breaking of rotational invariance. In that case, anomalies could be the imprints of a space-time anisotropy existing prior to inflation~\cite{Ackerman:2007nb,Gumrukcuoglu:2007bx}.

Theory discussing the evolution of CMB power spectra in an anisotropic inflation \cite{Pullen:2007tu,Pitrou:2008gk,Gessey-Jones:2021yky,Chang:2018msh} 
can presently be compared with just our one single universe, additionally constrained to small residual asymmetries. 
We show that both limitations can be overcome in analogue gravity experiments \cite{visser:review} with Bose-Einstein condensates (BEC) of particles with permanent dipoles \cite{PhysRevLett.94.160401,Dauxois2002}.

Analogue gravity \cite{visser:review} evolved from Unruh's seminal discovery of an analogue Hawking effect \cite{Hawking:1974rv} 
 in a transsonic fluid flow \cite{unruh:bholes}, arising since quantum sound waves propagate in an effective metric determined by the flow profile.
The latter can give rise to the sonic analog of a black hole event horizon, which has been realised and extensively studied in BEC \cite{PhysRevD.107.L121502,Garay:1999sk,Carusotto:2008ep,Lahav:2009wx,Larre:2011mq,Steinhauer:2014dra,steinhauer2,Boiron:2014npa,Steinhauer:2015saa,MunozdeNova:2018fxv,Isoard:2019buh,Kolobov:2019qfs,wuester:horizon,wuester:phonons,palan:losshawking:PhysRevA,Leonhardt:questioning_steinhauer_paper,Steinhauer2021,Wang2017,deNova:bhlaser_instability,Tettamanti:study,Isoard:departing_from_thermality_for_AHR,Tettamanti_nonthermalAHR}. Similarly, rotating BEC can furnish analogs of rotating Kerr black holes and the Penrose effect \cite{Solnyshkov:2018dgq,Kaur_hole_SOC_PhysRevA}, while expanding BEC or those with changing interaction strengths
can mimic expanding universes \cite{Eckel:2017uqx, fischer_analog,Jain:2007gg,barcelo:cpc,kurita_cpc_pra,Uhlmann_NJP_2005,Fischer_quantsim_inflation_PRA,Fedichev_cosmoprod_PRA,wuester:kerr} for the study of quantum fields during cosmological inflation.

Dipolar condensates have been shown to enhance entanglement of phonons created through the dynamical Casimir effect \cite{Tian_Rotonentang_PhysRevA}, allow studies of the impact of trans-Planckian modes on black-hole radiation \cite{PhysRevD.107.L121502} and the interplay between dispersion relations and scale invariance of power spectra following inflation \cite{fischer_analog}. However, only isotropic expanding universes were explored in analogue gravity so far \cite{barcelo:cpc,kurita_cpc_pra,Uhlmann_NJP_2005,Fischer_quantsim_inflation_PRA,Fedichev_cosmoprod_PRA}.
Our proposal will overcome this limitation, and thus provide the field of cosmology in anisotropic spacetimes with tuneable experiments
to study power spectra after complex inflation sequences, probe the effect of high frequency dispersion \cite{corley:moddispersion}, initial vacua \cite{GJ_inicond_PhysRevD,Kim_scalarfield_PhysRevD}
conversion of inhomogeneities into anisotropies \cite{Carroll:2008br} or instabilities \cite{Himmetoglu:2008zp,Himmetoglu:2008hx}. The dipolar BEC platform will also 
 enable interdisciplinary exchange with condensed matter and atomic physics communities \cite{Jacquet_nextgenexp}, exploring for example vacuum squeezing \cite{calz:hu,calz:hu2,wuester:kerr}.

\begin{figure}[htb]
	\centering
	\includegraphics[width=0.85\columnwidth,right]{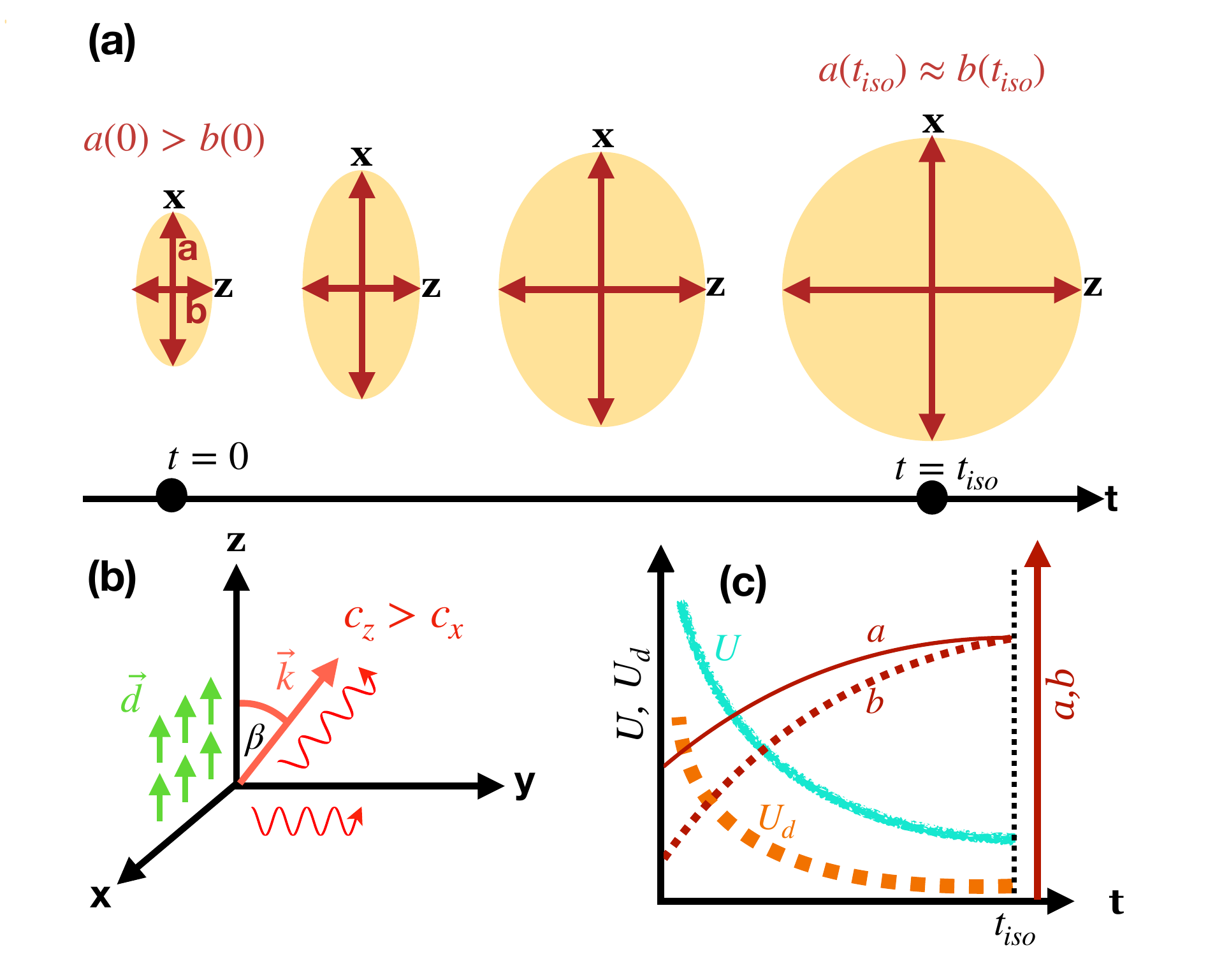} 
	\caption{\label{sketch} Inflation, with spacetime turning isotropic. (a) Scale factors $a(t)$, $b(t)$ for two orthogonal spatial dimensions, evolving from anisotropy to isotropy while expanding in laboratory time $t$.
		(b) In a dipolar BEC, the propagation speed $c$ of a phonon with wavevector $\mathbf{k}$ depends on the angle $\beta$ with the dipole direction $\mathbf{d}$.
		 (c) Scale factors are controlled via the strength of contact interactions, $U(t)$ and dipolar ones, $U_{d}(t)$, until the metric is isotropic at ${t}_{iso}$. 
	}
\end{figure}
%
\section{Foundation} 
In dipolar BEC, the speed of sound $c(\beta)$ depends on the angle ${\beta}$ between propagation direction of phonons and the dipolar axis $\mathbf{d}$ of the condensate atoms, see \frefp{sketch}{b}. In the gravitational analogy, this implies that the metric governing the propagation of sound waves acquires a preferred direction. 
In BEC this analogue metric can then be tuned from anisotropy to isotropy by control over the contact and dipolar interaction strengths. This exploits Feshbach resonances \cite{feshbach1,feshbach2}, to adjust the relative strength of s-wave and dipolar interactions \cite{strong_dipolar_lu,strong_dipolar_lahaye,dipolar_strength_lahaye} and time-averaged control of the dipolar interaction strength by rapidly rotating external fields \cite{dipolar_tuning_tang,dipolar_tuning_baillie,dipolar_tuning_giovanazzi}. Using both, the direction and degree of anisotropy can be temporally controlled in experiments.

In cosmology, anisotropies prior to inflation would impact the evolution of primordial density fluctuations in the inflaton field $\delta({\bf k})$ \cite{Ackerman:2007nb}, leading to residual signatures in their power spectrum defined through
$\langle \delta({\bf k})\delta^*({\bf q})\rangle =P(\bf k)\delta^3({\bf k}-{\bf q})$. Here $\bf k$, $\bf q$ are wave vectors of fluctuating modes. A violation of rotational invariance during the inflationary era can modify the power spectrum from an isotropic form $P(\mathbf{k})=P(k)$ to an anisotropic one:
\begin{equation} \label{eqn:aniso_power_spec}
    P'(\mathbf{k}) = P(k) + ({\bf \hat{k}}\cdot{\bf \hat{n}})^2 \Delta P(k),
\end{equation}
where ${\bf \hat{n}}$ is a unit vector along a preferred direction, $\hat{k}= \mathbf{k}/|\mathbf{k}|$ \cite{Ackerman:2007nb}, and $\Delta P(k)$ the amplitude of the anisotropic component.

In our analog universe, made from an expanding dipolar BEC, the power spectrum of phonon vacuum fluctuations also starts anisotropically, and can then be experimentally followed through its evolution 
while the universe expands and becomes isotropic. To demonstrate this, we tackle the initial phase of an inflation with direction dependent expansion rates as sketched in \fref{sketch}, analytically and through simulations, focussing on the retention of anisotropy in fluctuation spectra even at the time where the universe itself has become isotropic.

\section{Anisotropic effective space-time for phonons} 
%
The Hamiltonian for a dipolar BEC with atoms of mass $m$ is \cite{PhysRevA.63.053607,Lahaye_2009}
\begin{equation} \label{DBEC_Hamiltonian}
\hat{H}= \int{d^3{\bf{r}}\:\hat{\Psi}^\dagger(\mathbf{r},t)}\bigg[-\frac{\hbar^2\boldsymbol{\nabla}^2}{2m} +\frac{\hat{\phi}_{int}({\mathbf{r}},t)}{2}\bigg]\hat{\Psi}(\mathbf{r},t),
\end{equation}
with interaction operator 
\begin{equation}
\hat{\phi}_{int}({\bf{r}},t)=\int{d^3{\bf{r'}}\:\hat{\Psi}^\dagger({\bf{r'}},t)V_{int}({\bf{r}}-{\bf{r'}},t)\hat{\Psi}({\bf{r'}},t)},
\end{equation}
where $V_{int}(\mathbf{r}-\mathbf{r'},t)=U(t)\delta^{(3)}(\mathbf{r}-\mathbf{r'})+U_{\rm dd}(\mathbf{r}-\mathbf{r'},t)$ includes contact interactions of strength $U(t)$ and long-range dipole-dipole interactions (DDI) $U_{\rm {dd}}$. For $\psi=\expec{\hat{\Psi}}$, the mean field approximation of Heisenberg's equation,
known as Gross-Pitaevskii equation (GPE), is
\begin{equation}
\label{eqn:GPE}
i\hbar\frac{\partial\psi}{\partial t}=-\frac{\hbar^2}{2m}\boldsymbol{\nabla}^2\psi+\left(U(t)|\psi|^2+\fdd(t) \right)\psi,
\end{equation}
with 
\begin{equation}
\fdd(\mathbf{r},t)=\int|\psi(\mathbf{r},t)|^2\,U_{\rm dd}(\mathbf{r}-\mathbf{r'},t)\,{\rm d}^3\mathbf{r}'.
\end{equation}

Using the convolution theorem, the DDI can be expressed as $\fdd(\mathbf{r},t)={\cal F}^{-1}[\tilde{U}_{\rm dd}({\bds{k}},t)\tilde{n}(\bds{k},t)]$, where ${\cal F}$ denotes a Fourier transform, $\tilde{n}(\bds{k},t)={\cal F}[|\psi(\mathbf{r},t)|^2]$ and
\begin{equation}
\label{eqn:fdd}
\tilde{U}_{\rm dd}({\bds{k}},t)=U_d(t)(\cos^2\beta(\mathbf{k})-1/3)
\end{equation}
the dipole-dipole interaction in Fourier space. Writing $U_d(t)=\mu_0\mu(t)^2$, with $\mu_0$ the vacuum magnetic permeability, the dipole moment $\mu(t)$ of the atoms \cite{Lahaye_2009} is assumed
adjustable through external field averaging \cite{dipolar_tuning_tang,dipolar_tuning_baillie,dipolar_tuning_giovanazzi}. Here,
 $\beta$ is the angle between excitation wavenumber $\bds{k}$ and the constant polarization direction $\mathbf{d}$, which we take as our $z$-axis.
The contact interaction strength $U(t) = 4\pi\hbar^2a_s(t)/m$ is governed by the scattering length $a_s(t)$, which can also be varied in time using Feshbach resonances \cite{feshbach1,feshbach2}.

Expressing the condensate wavefunction as $\psi(\mathbf{r},t)=\sqrt{n(\mathbf{r},t)}e^{i\theta(\mathbf{r},t)}$ in \eref{eqn:GPE}, we obtain two coupled partial differential equations
\begin{eqnarray}
 \label{maineqn:nt_evo_eqn1}
\frac{\partial n}{\partial t}=-\frac{\hbar}{m}\Big[(\boldsymbol{\nabla} n)\cdot(\boldsymbol{\nabla}\theta)+n\boldsymbol{\nabla}^{2}\theta\Big],  \\
\frac{\partial\theta}{\partial t}=-\frac{\hbar}{2m}(\boldsymbol{\nabla}\theta)^{2}-\frac{Un}{\hbar} \quad-\frac{{U_d}}{\hbar}\mathcal{F}^{-1}\Big[f(\mathbf{k}) \tilde{n}\Big],
 \label{maineqn:nt_evo_eqn2}
\end{eqnarray}
 for real variables, density $n(\mathbf{r},t)$ and phase $\theta(\mathbf{r},t)$. We then re-instate small fluctuations on top of the mean field as $n\rightarrow n_{0}+\hat{n}_{1}$ and $\theta\rightarrow \theta_{0}+\hat{\theta}_{1}$, where $\hat{n}_{1}$ and $\hat{\theta}_{1}$ are the fluctuations and $n_{0}$ and $\theta_{0}$ are the background density and phase, respectively. Linearizing in $\hat{n}_{1}$ and $\hat{\theta}_{1}$, we can eliminate $\hat{n}_{1}$ as discussed in \ref{supplemental_derivation1}, to obtain an equation for phase fluctuations $\hat{\theta}_1$ of the form 
\begin{eqnarray}\label{eqn:field_equation}
\frac{1}{\sqrt{-g}} \; \partial_{\mu} \left( \sqrt{-g} \; g^{\mu \nu} \; \partial_{\nu} \hat{\theta}_1 \right) = 0,
\end{eqnarray}
defining an effective anisotropic metric tensor $g_{\mu\nu}$ with 
\begin{equation}
g_{\mu\mu} = \frac{n_0}{mc(t)}[-c^2(t), {\bar a}^2(t), {\bar a}^2(t), {\bar b}^2(t)]
\label{metric}
\end{equation}
on the diagonal, and $g_{\mu\nu}=0$ for $\mu\neq \nu$. Here $c(t)=\sqrt{n_0U(t)/m}$ is a fictitious speed of sound ignoring dipole interactions, while scale factors ${\bar a}(t)=[1-U_d(t)/3U(t)]^{-1/2}$ and ${\bar b}(t)=[1+2U_d(t)/3U(t)]^{-1/2}$ now incorporate the direction dependence of the true sound speed. We assumed a constant background density $n_0$, no condensate flow and dominant contact interactions $U_d(t)/3U(t) < 1$, refer to \ref{supplemental_derivation1}.  
Inflation in \eref{metric} shall arise dominantly through the time-dependence of contact interactions $U(t)=U_0f(t)$, where $U_{0}$ is the interaction strength at $t=0$ and $f(t)$ specified later. Meanwhile the relative importance of dipolar interactions governs (an)isotropy. Defining $c_0^2=n_0U_0/m$, the line element in the laboratory frame can then be written as 
\begin{equation}
\label{metric_labtime}
ds^2 = -c_0^2\sqrt{f(t)} dt^2 + \frac{{\bar a}^2(t)}{\sqrt{f(t)}} (dx^2+dy^2)+\frac{{\bar b}^2(t)}{\sqrt{f(t)}} dz^2.
\end{equation}
To see the analogy to cosmology more clearly, we employ the time transformation $d\eta^2=\sqrt{f(t)} dt^2$ to reach
\begin{equation}\label{conf_aniso_metric}
ds^2 = -c_0^2d\eta^2 + a^2(\eta) (dx^2+dy^2)+b^2(\eta) dz^2.
\end{equation}
with $a^2(\eta)={{\bar a}^2(\eta)}/{\sqrt{f(\eta)}}$  and $b^2(\eta)={{\bar b}^2(\eta)}/{\sqrt{f(\eta)}}$.
Now, we construct an anisotropically expanding analogue inflationary universe, which evolves into an isotropic one and calculate the expected phonon fluctuation power spectrum, starting from an initial vacuum state. For this, we chose $a(\eta)=a_0e^{H_a\eta}$ and $b(\eta)=b_0e^{H_b\eta}$, with two different (constant) Hubble parameters $H_a=\dot{a}(\eta)/a(\eta)$ and $H_b=\dot{b}(\eta)/b(\eta)$. 
We will also refer to the average Hubble parameter $\bar{H}=(2H_a+H_b)/3$ and deviation from dynamic isotropy as  $\epsilon_H=2(H_b-H_a)/3\bar{H}$.
Together, our ansatz $U(t)=U_0f(t)$ for the time variation of s-wave interactions and the target evolution of anisotropic scale factors, $a(\eta)$ and $b(\eta)$, now fix the relation between conformal time and laboratory time and required form of $U_{d}(t)=\mu_0\mu_m^2 h(t)/4\pi$ with $h(0)=1$, as shown in \ref{supplemental_derivation2}.

\section{Power spectrum of fluctuation correlations}
%
A key observable that can record the imprint of a possible anisotropy in the early universe is the fluctuation power spectrum, the analogue of which we propose to experimentally probe in tuneable experiments with dipolar BEC. 
Here we define the power spectrum through $P(\kvec)=\langle \hat{a}^{\dagger}_{\kvec} \hat{a}_{\kvec}\rangle$,
as vacuum expectation value of plane wave modes of the phase fluctuation field 
\begin{equation} \label{eq:FourierDef}
	\hat{\theta}_1(\xvec,t) = \int\frac{d^{3}\kvec}{(2\pi)^3}\left(e^{i \kvec \cdot \xvec} \tilde{\theta}_1(\kvec,t) \hat{a}_{\kvec} 
	+ e^{-i \kvec \cdot \xvec} \tilde{\theta}^*_1(\kvec,t) \hat{a}^{\dagger}_{\kvec}\right)\, ,
\end{equation}
quantized here as usual by expanding in Fourier modes. The creation and annihilation operators $\hat{a}^{\dagger}_{\kvec}$ and $\hat{a}_{\kvec}$ satisfy the standard Bosonic commutation
relations.

The power spectrum can be found via 
\begin{equation}
	\int \frac{d^3\kvec}{(2\pi)^3}e^{-i \kvec\cdot (\mathbf{r} -\mathbf{r'})}P(\kvec) = \langle0|\hat{\theta}_1(\mathbf{r},t)\hat{\theta}_1(\mathbf{r'},t)|0\rangle
\end{equation}
as Fourier transform of the phase correlation function. Since we are considering a homogenous system, 
the latter can only depend on the relative coordinate $\mathbf{r} -\mathbf{r'}$.

Condensate phase correlations can be measured through interference experiments \cite{Hellweg_measurephasecorrel_PhysRevLett,Cacciapuoti_measurephasecorrel_PhysRevA}, or phase fluctuations could first be related to density fluctuations \cite{Fischer_backreaction}. 
Then high resolution density-density correlations can be recorded in experiments \cite{Steinhauer:2015ava,Steinhauer:2014dra,Steinhauer:2015saa}.

\begin{figure}[htb]
\includegraphics[width=0.85\columnwidth,right]{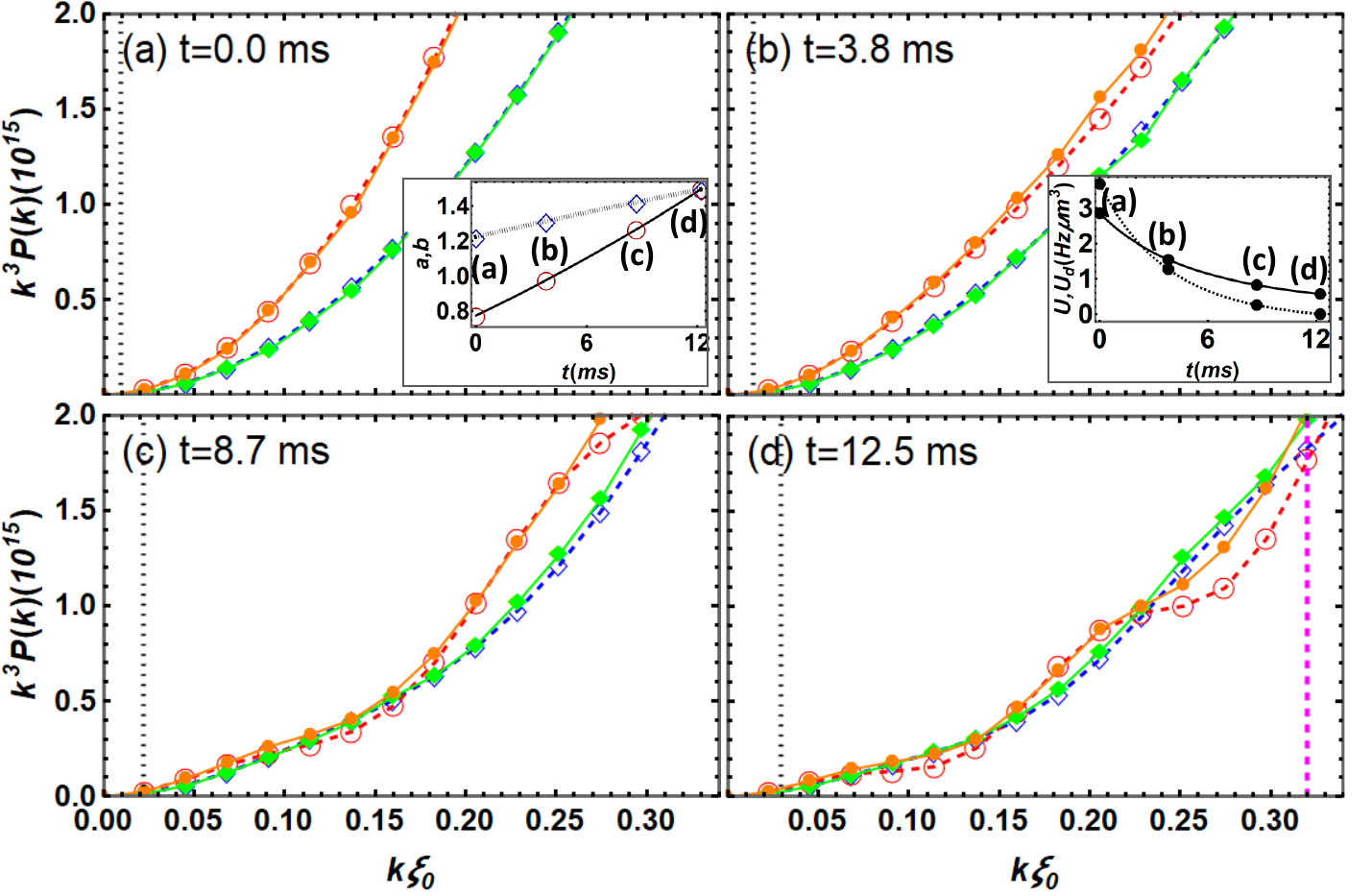}
\caption
{
\label{power_spectra_fig} 
Phonon power-spectra during analogue inflation, at lab times $t$ indicated. We compare analytical spectra from \eref{bec_fluc_eqn} (dashed, empty symbols) with numerical TWA simulations (solid, filled symbols), for wavenumbers $k_z$ along the $z$-axis (solid orange, red dashed), and $k_x$ along the $x$-axis (solid green, blue dashed), which are scaled with the \emph{initial} healing length from contact interactions $\xi_0$. The inset in (a) shows the scale factors $a(\eta(t))$, $b(\eta(t))$ and the one in (b) the contact interaction strength $U(t)$ and dipolar strength $U_d(t)$.
Vertical dotted black lines indicate the the Hubble wavenumber $K_h=1/R_h$ where $R_h=\bar{c}(t)/\bar{H}$ with $\bar{c}(t)=\sqrt{n_0 (U(t)+U_d(t))/m}$, and the vertical magenta dashed line the time evolving inverse healing length $k_\xi=\xi(t)^{-1}=\sqrt{2mn_0(U(t)+U_d(t))}/\hbar$. See also supplementary movies.
}
\end{figure}

Inserting $\hat{\theta}_1$ into \eref{eqn:field_equation}, the metric \bref{metric_labtime} implies  
\begin{equation}\label{bec_fluc_eqn}
\frac{\partial^2\tilde{\theta}_{1}}{\partial t^2}+{\gamma(t)} \frac{\partial\tilde{\theta}_{1}}{\partial t}
+\omega(t)^2\tilde{\theta}_{1} = 0,
\end{equation}
the equation of motion of a damped harmonic oscillator with time-dependent frequency 
$\omega(t)=(\mathbf{k}^2 n_0 \mathcal{Q}/m)^{1/2}$ and damping rate $\gamma(t)= \mathcal{Q}(\partial \mathcal{Q}/\partial t)$,
using $\mathcal{Q}=\mathcal{Q}(\mathbf{k},t)=-U(t)-{U}_d(t)\Big[\cos^{2}\beta(\mathbf{k})-(1/3)\Big]$. 
We convert \eref{bec_fluc_eqn} into the equivalent Hamilton equations,
\begin{equation} \label{hamiltonian_eoms}
\dot{q}(t)=p(t), \hspace{0.5cm}  \dot{p}(t)=-\gamma(t)p(t)-\omega(\mathbf{k},t)^2q(t),
\end{equation}
from which we can construct complex mode amplitudes $\tilde{\theta}_1(\kvec,t)=q(\mathbf{k} ,t)+i p(\mathbf{k} ,t)/\omega(t)$.
In the vacuum $\hat{a}_{\kvec} \ket{0}=0$, we then have
$P(\mathbf{k},t)=|\tilde{\theta}_1(\kvec,t)|^2$. We solve \eref{hamiltonian_eoms} numerically with initial conditions $p(0)=0$ and $\tilde{\theta}(0)$ matched onto the Bogoliubov vacuum of the initial state of the dipolar BEC, discussed in \ref{app:twa_deriv}.
Finding $P(\mathbf{k},t)=|\tilde{\theta}_1(\mathbf{k} ,t)|^2$, using the
initial conditions obtained from the Bogoliubov equations of the initial dipolar BEC, the resultant $k^3P(k)$ is shown in \fref{power_spectra_fig}.

For the demonstration, we assume a dipolar BEC of Erbium atoms \cite{erbium1, Chomaz_magnetig_gases_review}, each of mass $m=2.8\times10^{-25}$ kg, with initial magnetic dipolar moment ${\mu_m} = 1.897\mu_B$ already reduced compared to the usual $\bar{\mu}_m = 7\mu_B$, where $\mu_B$ is the Bohr magneton.
The initial modified s-wave scattering length is $a_s= 0.599$ nm and homogenous density $n_0=5\times10^{20}$ m$^{-3}$, yielding an initial healing length $\xi_0=0.364\mu$m. The inflationary parameters are taken as $a_0=1.225$, $b_0=0.775$, $H_a=(200/q) s^{-1}$ and $H_b=(658/q)s^{-1} $, where the factor $q$ just scales the expansion rate.
For these choices, the metric becomes isotropic at a lab time ${t}_{iso}=q\times 1.25$ ms. 

The evolving power spectrum thus obtained is shown in \fref{power_spectra_fig} for the case of $q=10$. Fourier components of correlations in different directions have different strength initially, a signature of an anisotropic Bogoliubov vacuum. As the analogue universe expands, it also becomes more isotropic, since the two scale factors approach each other. Consequently the power spectrum changes from strongly anisotropic to nearly isotropic. At
 $t={t}_{iso}$, shown in panel \fref{power_spectra_fig}(d), small imprints of the initial anisotropy still remain, although the metric has become isotropic. 
Experiments could naturally handle much more extreme inflation sequences than the one here, and probe additional topics actively explored in cosmology, such as unstable modes \cite{Himmetoglu:2008zp,Himmetoglu:2008hx} and the conversion of inhomogeneity into anisotropy \cite{Carroll:2008br}.

\section{Beyond mean field simulations}
%
To confirm the analogue model, we numerically simulate the same inflation with the Truncated Wigner Approximation (TWA) 
\cite{twa_isella,twa_norrie,twa_sinatra,twa_sinatra_prl,Davis_2013,twa_wuster,PhysRevA.58.4824}, which can provide the quantum field evolution from \eref{DBEC_Hamiltonian} as long as fluctuations remain small. Unlike the calculations based on the metric \bref{metric}, these simulations also describe BEC excitations with wavenumbers $k\xi(t)>1$ for which the analogy does not hold. They further would cover particle creation \cite{Jain:2007gg}, which is absent here, the interaction of quasiparticles, and can verify the dynamical stability of the mean field background on time-scales of interest.

In TWA, one generates an ensemble of stochastic fluctuations added to the mean field, to sample the Wigner quasi-distribution function of the initial density operator. The quantum field dynamics is then found from noisy GPE simulations. We extract the power spectrum from phase fluctuation correlation functions via
$P({\bf{k}},t)=\int d^{3}\mathbf{r}_0  \int d^{3}\mathbf{r'} \langle \hat{\theta}_{1}(\mathbf{r}_0,t)\hat{\theta}_{1}(\mathbf{r}_0+\mathbf{r'},t)\rangle  e^{-i \mathbf{k}  \cdot \mathbf{r'} } /V$,
 as discussed in the \ref{app:twa_deriv}, and use the same parameters as before, in a cubic box of volume $V=(50{\mu}m)^3$ with $(64)^3$ gridpoints and ${N}_{traj}=5120$ stochastic trajectories.
TWA power spectra confirm our analytical results, as shown in \fref{power_spectra_fig}, and thus verify that there is no disturbing effect of 
single particle excitations at high wavenumbers and that dynamic instabilities of the mean field are absent. These would only occur in dipolar BEC for larger dipolar interaction strength \cite{PhysRevLett.85.1791,PhysRevA.71.033618,t.koch1,PhysRevLett.101.080401}. 

\begin{figure}[htb]
\includegraphics[width=0.85\columnwidth,right]{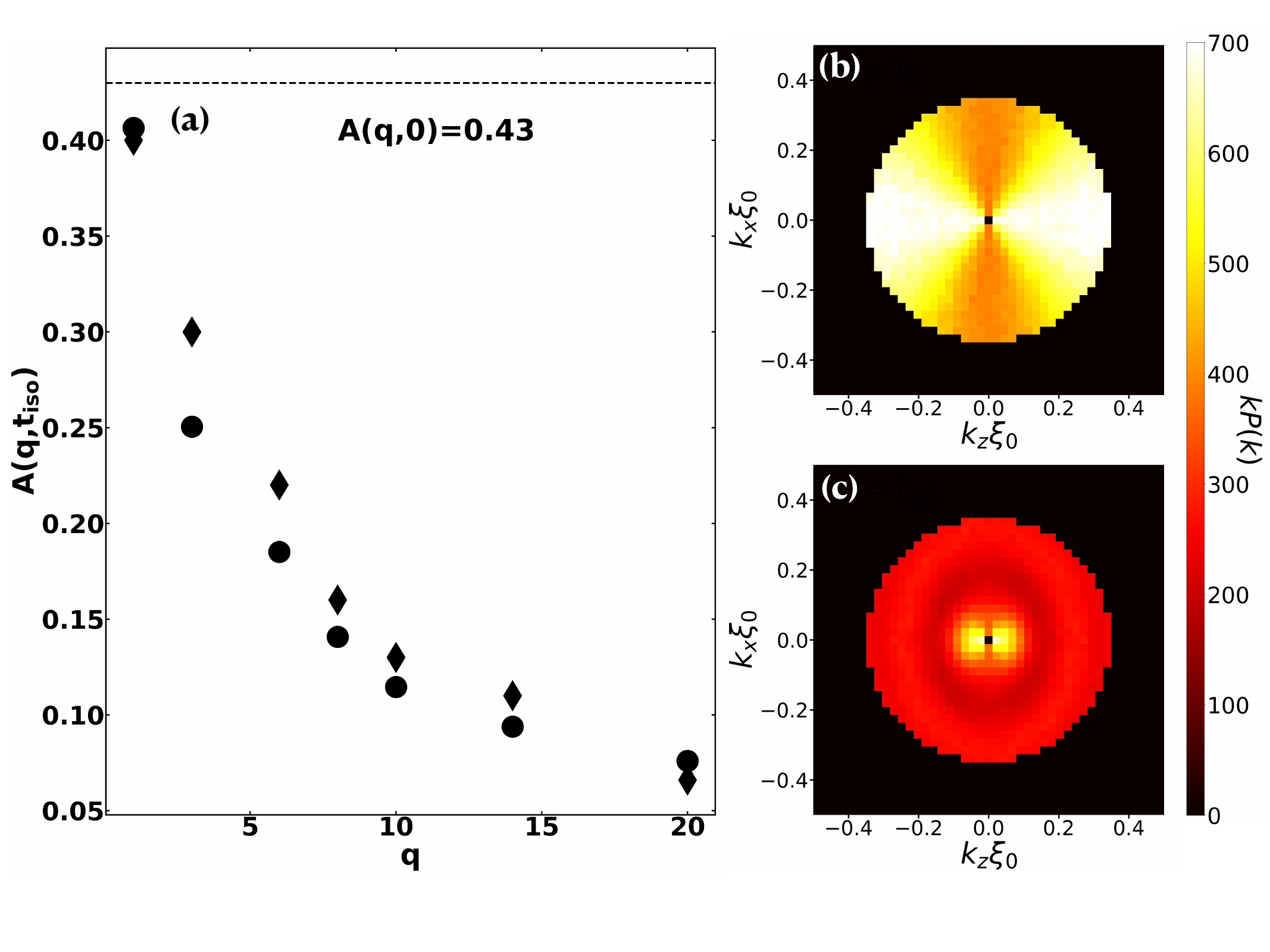}
\caption{\label{power_spectra_results} (a) Isotropisation in $A(q,{t}_{iso})$ at ${t}_{iso}$ for different inflation rates $\sim q^{-1}$, from \eref{bec_fluc_eqn} (black $\bullet$) and TWA (black $\blacklozenge$). The  horizontal dashed line is the initial anisotropy $A(q,0)$. (b,c) show power spectra $|\mathbf{k}| P(\mathbf{k})$ at the initial time, $t=0$ (b), and final time, $t={t}_{iso}$ (c), in the $(k_x,k_z)$ plane from TWA simulations.
}
\end{figure}

Analog gravity has thus allowed us to map isotropisation during cosmic inflation to continuous variations of a many-body Hamiltonian.
The slower the Hamiltonian changes, the better the system will be able to adiabatically follow the quantum ground-state. The latter will be isotropic for an isotropic system, unless there is spontaneous symmetry breaking. We thus expect final power spectra to be more isotropic at $t={t}_{iso}$ for slow evolution (large $q$). This is indeed what we find, as shown in \fref{power_spectra_results}. We have defined the net anisotropy of a spectrum as $A(q,t)=[\bar{P}_z-\bar{P}_x]/\bar{P}_z$, with $\bar{P}_j=\int_{0}^{{k}_{max}} dk_j \:k_j  P(k_j,t)$, where the upper integration limit is the largest wavenumber containing noise in TWA, ${k}_{max}=0.94$ $\mu$m$^{-1}$ for \fref{power_spectra_results}. The figure also shows more detailed cuts through power-spectra from TWA in the $(k_x,k_z)$ plane, illustrating that $|\mathbf{k}| P$ only depends on $\beta$ initially (see \ref{app:twa_deriv}), which is why we have chosen it as integrand for $A(q,t)$. During inflation, the function $|\mathbf{k}| P$ then acquires nontrivial structure, shown in \frefp{power_spectra_results}{c}.

An important dynamical scale during cosmic inflation is the Hubble radius $R_h(t) = \bar{c}(t)/\bar{H}\sim q$ (Hubble wavenumber $K_h=R_h^{-1}$). Only modes with wavelengths $\lambda<R_h(t)$ will be oscillating, while those with $\lambda>R_h(t)$ freeze out \cite{dodelson:2003}. The latter are situated on the left of the vertical blue dotted lines in \fref{power_spectra_fig}, but would contain most modes shown for the lower $q$. Meanwhile, the analog metric \bref{metric_labtime} only describes long wavelength modes with $k \xi(t)< 1$, to the left of the magenta dashed vertical line in \frefp{power_spectra_fig}{d}, and at larger $k$ in other panels. We thus demonstrated that one can study both, frozen and unfrozen modes, with wavenumbers for which the analogy is valid. Dipolar BEC can have lifetimes of a few hundred milliseconds even while tuning interactions \cite{dipolar_tuning_tang,dipolar_tuning_baillie}, and all chosen isotropisation times ${t}_{iso}$ are shorter.

\section{Conclusions and outlook}
%
We have shown that dipolar Bose-Einstein condensates can provide an experimental window on the dynamics of quantum fields during anisotropic cosmological inflation, which was
was hitherto experimentally inaccessible, except for observations of our one single universe. Thus one can probe different residual anisotropies after a given inflation sequence, conversion of inhomogeneities into anisotropies \cite{Carroll:2008br}, instabilities \cite{Himmetoglu:2008zp,Himmetoglu:2008hx}
or mode squeezing \cite{calz:hu,calz:hu2,wuester:kerr}. If the condensate is given a finite flow velocity, the same experimental platform can also create analogue black holes in anisotropic space times. By tuning the initial fluctuations, we can explore the analog of primoridal gravitational waves and how these would later reflect an initial anisotropy of the universe, motivated by \rref{Ito:2016aai} predicting that the detection of gravitational wave in the 10-100 MHz regime would solidify the occurrence of anisotropic inflation. Instead of dipolar BEC, anisotropic analogue space times could also be engineered using spin-orbit coupling \cite{PhysRevA.90.023627,Kumar_2018,Kaur_hole_SOC_PhysRevA,PhysRevA.102.023314}, and even more tunability might arise from combining the two. 

\section*{Data availability statement}
The data that support the findings of this study are available upon request from the authors.
\ack
We gratefully acknowledge financial support from the Max-Planck society under the MPG-India partner group program
and helpful comments from Rejish Nath. S.P. would like to thank DST (Govt. of India) for the financial support under Grant No. SERB/PHY/2021057.
%
\appendix
\setcounter{secnumdepth}{2}
\section{Derivation of the anisotropic metric}
\label{supplemental_derivation1}
\setcounter{equation}{0}
To investigate dipolar BEC, consider the Gross-Pitaevskii (GP) equation for the $3+1$-D case as 
\begin{equation}
\label{app:GPE}
i\hbar\frac{\partial\psi}{\partial t}=-\frac{\hbar^2}{2m}\boldsymbol{\nabla}^2\psi+\left(V_{\rm ext}+U|\psi|^2+\fdd\right)\psi,
\end{equation}
where $\fdd$ is the dipolar mean field interaction
\begin{equation}
\fdd(\mathbf{r},t)=\int|\psi(\mathbf{r'},t)|^2\,U_{\rm dd}(\mathbf{r}-\mathbf{r'})\,{\rm d}^3\mathbf{r'}.
\end{equation}
Using the convolution theorem, \eref{app:GPE} becomes 
\begin{eqnarray}
&i\hbar \frac{\partial \psi(\mathbf{r},t)}{\partial t}=-\frac{\hbar^{2}}{2m} \boldsymbol{\nabla}^{2}\psi(\mathbf{r},t) \nonumber
+U|\psi(\mathbf{r},t)|^{2}\psi(\mathbf{r},t)\\ 
&+U_{d}\Big(\int{\frac{d^3\mathbf{k}}{(2\pi)^{3}}} e^{i\kvec \cdot \xvec} f(\mathbf{k}) 
\tilde{n}(\mathbf{k},t)\Big)\psi(\mathbf{r},t).
\label{GPE_explicit}
\end{eqnarray}

Here, $U=4\pi\hbar^{2}a_{s}/m$ where $a_{s}$ is the s-wave scattering length and $m$ is the mass of particles constituting the BEC. The dipolar interaction strength is $U_{d}=\mu_{0}\mu^{2}$ where $\mu$ is the dipole moment of the BEC particles and $\mu_0$ the permeability of the vacuum. The function $\tilde{n}(\mathbf{k},t)$ denotes the Fourier transform of the atomic number density $|\psi(\mathbf{r},t)|^{2}$. The interaction kernel $f(\mathbf{k})$ is given by
\begin{equation}
f(\mathbf{k})=\frac{3(\hat{\mathbf{k}}\cdot\hat{\mathbf{d}})^2-1}{3}=\frac{3\cos^{2}\beta-1}{3},
\label{fkernel}
\end{equation}
where $\beta$ is the angle between the wavevector direction $\hat{\mathbf{k}}=\mathbf{k}/|\mathbf{k}|$ and the dipole axis $\hat{\mathbf{d}}=\mathbf{d}/|\mathbf{d}|$, which we choose to define the z-axis and keep constant.
Now, to obtain the metric from \eref{GPE_explicit}, we use the Madelung ansatz for the wavefunction $\psi(\mathbf{r},t)=\sqrt{n(\mathbf{r},t)} e^{i\theta(\mathbf{r},t)}$ to derive evolution equations for $n(\mathbf{r},t)$ and $\theta(\mathbf{r},t)$ as
%
\begin{eqnarray}
 \label{eqn:nt_evo_eqn1}
&\frac{\partial n}{\partial t}=-\frac{\hbar}{m}\Big[(\boldsymbol{\nabla} n)\cdot(\boldsymbol{\nabla}\theta)+n\boldsymbol{\nabla}^{2}\theta\Big],  \\
&\frac{\partial\theta}{\partial t}=-\frac{\hbar}{2m}(\boldsymbol{\nabla}\theta)^{2}-\frac{Un}{\hbar} \quad-\frac{{U_d}}{\hbar}\mathcal{F}^{-1}\Big[f(\mathbf{k}) \tilde{n}\Big],
 \label{eqn:nt_evo_eqn2}
\end{eqnarray}
%
where $\mathcal{F}^{-1}$ denotes the inverse Fourier transform and we have omitted the arguments of $n$ and $\theta$ for the purpose of compactness.

\par 
Next, we wish to obtain equation for fluctuations about the mean field and replace $n$ and $\theta$ as $n\rightarrow n_{0}+n_{1}$ and $\theta\rightarrow \theta_{0}+\theta_{1}$, where $n_{1}$ and $\theta_{1}$ are small amplitude fluctuations. We thus focus on fluctuations around the mean field $\sqrt{n_{0}}e^{i\theta_{0}}$. Further, we assume that the mean field has no flow velocity associated with it, $\boldsymbol{\nabla}\theta_{0}=0$, and that the mean density is constant over space, $\boldsymbol{\nabla} n_{0}=0$. Both are well satisfied near the centre of a large BEC in the Thomas-Fermi limit. With these assumptions and linearization in the small amplitude fields $n_{1}$ and $\theta_{1}$, we turn \esref{eqn:nt_evo_eqn1}{eqn:nt_evo_eqn2} into
%
\begin{eqnarray}
&\frac{\partial n_{1}}{\partial t}=-\frac{\hbar}{m}\Big[n_{0}\boldsymbol{\nabla}^{2}\theta_{1}\Big],\\
&\frac{\partial\theta_{1}}{\partial t}=-\frac{Un_{1}}{\hbar}-\frac{U_d}{\hbar}\mathcal{F}^{-1}\Big[f(\mathbf{k}) \tilde{n}_{1}\Big].
\end{eqnarray}
%
Taking the Fourier transform w.r.t.~spatial dimensions of the above equation yields
%
\begin{eqnarray}
&\frac{\partial \tilde{n}_{1}}{\partial t}=\frac{\hbar}{m}\big[n_{0} (k_x^{2}+k_y^{2}+k_z^{2}) \tilde{\theta}_{1}\big],  
\label{dn1_dt}
\\
&\frac{\partial\tilde{\theta}_{1}}{\partial t}=-\frac{U\tilde{n}_{1}}{\hbar}-\frac{{U}_d}{\hbar}\Big[f(\mathbf{k}) \tilde{n}_{1}\Big],
\label{dtheta1_dt}
\end{eqnarray}
%
using the short hand $\tilde{\theta}_{1}$ and $\tilde{n}_{1}$ for Fourier space fluctuations.

We can formally solve \eref{dtheta1_dt} for
\begin{eqnarray}
\tilde{n}_{1}&=\frac{\partial\tilde{\theta}_{1}}{\partial t}\times\Big\{-\frac{ U}{\hbar}-\frac{{U}_d}{\hbar }\Big[f(\mathbf{k}) \Big] \Big\}^{-1},
\label{eq:inv1}
\end{eqnarray}
and insert this into \eref{dn1_dt}, using \eref{fkernel} to find: 
\begin{eqnarray}\label{eqn:comp_field_eqn}
\frac{\partial}{\partial t}\bigg(\frac{\partial\tilde{\theta}_{1}}{\partial t}\times\Big\{-U-{U}_d\Big[\frac{-k_x^2-k_y^2+ 2 k_z^2}{3\mathbf{k}^2} \Big] \Big\}^{-1}\bigg)
-\frac{n_{0}}{m}[ \mathbf{k}^2 \tilde{\theta}_{1}] = 0,
\end{eqnarray}
which is an equation for the phase fluctuations alone.
 
To obtain the metric, we compare \eref{eqn:comp_field_eqn} with the Fourier transform of \eref{eqn:field_equation}  and see 
%
\begin{equation*}
g_{\mu\nu} =\frac{n_0}{mc(t)}
\left[
\begin{array}{cccc}
-c_0^2 & 0 & 0 & 0 \\
0 & {\bar a}(t)^{2} & 0 & 0 \\
0 &  0 & {\bar a}(t)^{2}  & 0 \\
0 & 0 & 0 & {\bar b}(t)^{2}
\end{array}
\right]
\label{metric1}
\end{equation*}
with $ {\bar a}(t)$, ${\bar b}(t)$ defined as
\begin{equation}
\frac{1}{{\bar a}(t)^2}=\bigg(1-\frac{{U}_d(t)}{3 U(t)}\bigg),\quad \frac{1}{{\bar b}(t)^2}=\bigg(1+\frac{2 {U}_d(t)}{3 U(t)}\bigg).
\end{equation}
Whenever the dipole interactions are absent and thus ${U}_d=0$, the metric is isotropic as expected.
%
\section{Anisotropic analogue inflation in BEC}
\label{supplemental_derivation2}
%
We can rewrite the metric \bref{metric1}
using $c^2(t)=n_0U(t)/m$, inserting the parametrisation of time dependent contact interactions $U(t)=U_0f(t)$ and definitions $c_0^2=n_0U_0/m$ and $\Omega_0^2=\sqrt{\frac{n_0}{mU_0}}$ as
\begin{equation*}
g_{\mu\nu} = \Omega_0^2
\left[
\begin{array}{cccc}
-c_0^2\sqrt{f(t)} & 0 & 0 & 0 \\
0 & \frac{{\bar a}(t)^{2}}{\sqrt{f(t)}} & 0 & 0 \\
0 &  0 & \frac{{\bar a}(t)^{2}}{\sqrt{f(t)}}  & 0 \\
0 & 0 & 0 & \frac{{\bar b}(t)^{2}}{\sqrt{f(t)}}
\end{array}
\right]
\end{equation*}
from which we remove the conformal factor $ \Omega_0^2$ with the definition $g_{\mu\nu} = \Omega_0^2 \tilde{g}_{\mu\nu}$ and then express the
line element in terms of  $\tilde{g}_{\mu\nu}$
\begin{equation} \label{lab_frame_metric}
ds^2 = -c_0^2\sqrt{f(t)} dt^2 + \frac{{\bar a}(t)^{2}}{\sqrt{f(t)}} (dx^2+dy^2)+\frac{{\bar b}(t)^{2}}{\sqrt{f(t)}} dz^2 .
\end{equation}
Now, we re-define the time-coordinate as
\begin{equation} \label{time_transf}
d\eta^2=\sqrt{f(t)} dt^2,
\end{equation}
and write the line element in the new  coordinates as,
\begin{equation}
ds^2 = -c_0^2d\eta^2 + a^2(\eta) (dx^2+dy^2)+b^2(\eta) dz^2.
\end{equation}
For the analogue inflationary universe to expand anisotropically, we take $a(\eta)=a_0e^{H_a\eta}$ and $b(\eta)=b_0e^{H_b\eta}$ where $H_a=\dot{a}(\eta)/a(\eta)$ and $H_b=\dot{b}(\eta)/b(\eta)$, $U(t)=U_0f(t)$ and $U_d(t)=\mu_0\mu_m^2h(t)$, where $f(t)$ and $h(t)$ contain the time dependent part of contact and dipolar interactions respectively.
Using these relations, we reach
\begin{eqnarray} \label{def_ft}
f(t)=\bigg\{\big[(2/a_0^2) e^{-2H_a\eta(t)}+(1/b_0^2)e^{-2H_b\eta(t)}\big]/3\bigg\}^2 
\end{eqnarray}
and
\begin{eqnarray}
h(t)= {}& \big[(2/a_0^2) e^{-2H_a\eta(t)}+(1/b_0^2)e^{-2H_b\eta(t)}\big)] \label{def_ht}\nonumber\\
&\times \big[(-1/a_0^2) e^{-2H_a\eta(t)}+(1/b_0^2)e^{-2H_b\eta(t)}\big]/3.
\end{eqnarray}
Now using \eref{time_transf} and \eref{def_ft}, we find the relation between transformed time $\eta$ and lab time $t$, which we express in the form $\eta(t)=\sum_{j=1}^{l} c_j t^j$ where the coefficients $c_j$ depend on Hubble parameters $H_{a,b}$ and thus on the inflation rate control parameter $q$. This dependence arises since $f(t)$, $h(t)$ depend on $H_a$ and $H_b$, which in turn depend on $q$. From $\eta(t)$ we can insert \eref{def_ft} and \eref{def_ht} into $U(t)=U_0f(t)$ and $U_d(t)=\mu_0\mu_m^2h(t)$ to generate a target inflationary scenario.

We have now provided a complete recipe for tuning the interactions such that one obtains an anisotropically expanding universe in dipolar BEC.
The same recipe can also be used to implement a different functional form for scale factors in conformal time than the one assumed above.

\section{Truncated Wigner simulations}
\label{app:twa_deriv}
%
Here we describe how correlations of phase fluctuations can be obtained from TWA averages. We start from the Bose field operator written as a sum of mean field and quantum fluctuations. In the Madelung ansatz, $\hat{\Psi}({\mathbf{r}},t)= \sqrt{n_{0}+\hat{n}_{1}({\mathbf{r}},t)} \; e^{i(\theta_{0}+\hat{\theta}_{1}({\mathbf{r}},t))}$, where $\hat{n}_{1}$ and $\hat{\theta}_{1}$ represent density and phase fluctuations respectively. Assuming that fluctuations are small compared to the mean field, the field operator and consequently, the fluctuations may be written as
\begin{eqnarray}
\hat{\Psi}({\mathbf{r}},t)&=\Psi_{0}+\delta \hat{\Psi}({\mathbf{r}},t) \nonumber\\
&=\sqrt{n_{0}}+\sqrt{n_{0}}\Big(\frac{\hat{n}_{1}({\mathbf{r}},t)}{2n_{0}}+i\hat{\theta}_{1}({\mathbf{r}},t)\Big),\\
& \nonumber\\
\hat{n}_{1}({\mathbf{r}},t)&=\sqrt{n_{0}}\Big(\delta \hat{\Psi}({\mathbf{r}},t)+\delta \hat{\Psi}^{\dagger}({\mathbf{r}},t)\Big)\\
\hat{\theta}_{1}({\mathbf{r}},t)&=\frac{i}{2\sqrt{n_{0}}} \Big(-\delta \hat{\Psi}({\mathbf{r}},t)+\delta \hat{\Psi}^{\dagger}({\mathbf{r}},t)\Big) .
\label{eq:flucs1}
\end{eqnarray}
With this form of the phase fluctuations $\hat{\theta}_{1}$ we can write the phase correlations as
\begin{eqnarray}
&\langle\hat{\theta}_{1}({\mathbf{r}},t)\hat{\theta}_{1}({\mathbf{r}+\mathbf{r'}},t)\rangle= \frac{1}{4n_{0}} \Big(-\langle\delta \hat{\Psi}({\mathbf{r}},t)\delta \hat{\Psi}({\mathbf{r}+\mathbf{r'}},t)\rangle \CR
&+ \langle\delta \hat{\Psi}({\mathbf{r}},t)\delta \hat{\Psi}^{\dagger}({\mathbf{r}+\mathbf{r'}},t)\rangle+ \langle\delta \hat{\Psi}^{\dagger}({\mathbf{r}},t)\delta \hat{\Psi}({\mathbf{r}+\mathbf{r'}},t)\rangle \CR
&-\langle\delta \hat{\Psi}^{\dagger}({\mathbf{r}},t)\delta \hat{\Psi}^{\dagger}({\mathbf{r}+\mathbf{r'}},t)\rangle\Big).
\label{eq:phase_corr1}
\end{eqnarray}
We know that truncated Wigner averages $\expec{\cdots}_W$ provide an approximation for symmetrically ordered expectation values
of field operators:
\begin{eqnarray}
 &\langle \alpha^{*}({\mathbf{r}},t) \alpha({\bf{r'}},t) \rangle _{W} 
 =\left(\langle \hat{\Psi}^{\dagger}({\bf{r}},t) \hat{\Psi}({\bf r'},t) + \hat{\Psi}({\bf r'},t)\hat{\Psi}^{\dagger}({\bf{r}},t) \rangle\right)/2.
\label{eq:twa_avg1}
\end{eqnarray}
Hence, the correlation of phase fluctuations can be expressed in terms of TWA averages in position space as
\begin{eqnarray}
&\langle\hat{\theta}_{1}({\mathbf{r}},t)\hat{\theta}_{1}({\mathbf{r}+\mathbf{r'}},t)\rangle=
\frac{1}{4}\Big[- \frac{\langle\alpha({\mathbf{r}},t)\alpha({\mathbf{r}+\mathbf{r'}},t)\rangle _{W}}{\langle\alpha({\mathbf{r}},t) \rangle _{W}\langle\alpha({\mathbf{r}+\mathbf{r'}},t) \rangle _{W}}\CR
&+ \frac{\langle\alpha({\mathbf{r}},t)\alpha^{*}({\mathbf{r}+\mathbf{r'}},t)\rangle _{W}}{\langle\alpha({\mathbf{r}},t) \rangle _{W}\langle\alpha^{*}({\mathbf{r}+\mathbf{r'}},t) \rangle _{W}}
+\frac{\langle\alpha^{*}({\mathbf{r}},t)\alpha({\mathbf{r}+\mathbf{r'}},t)\rangle _{W}}{\langle\alpha^{*}({\mathbf{r}},t) \rangle _{W}\langle\alpha({\mathbf{r}+\mathbf{r'}},t) \rangle _{W}}\CR
&-\frac{\langle\alpha^{*}({\mathbf{r}},t)\alpha^{*}({\mathbf{r}+\mathbf{r'}},t)\rangle _{W}}{\langle\alpha^{*}({\mathbf{r}},t) \rangle _{W}\langle\alpha^{*}({\mathbf{r}+\mathbf{r'}},t) \rangle _{W}} \Big],
\label{eq:phase_corr2}
\end{eqnarray}
where $\langle\alpha({\mathbf{r}},t)\rangle _{W}= \sqrt{n_{0}}$. Since we consider a homogeneous system, these correlation do not depend on $\mathbf{r}$ and we average over that coordinate to increase statistics. 
The power spectrum using the $3D$ correlation function is written as 
\begin{equation} \label{eq:ps_sim}
P(\kvec,t)=\int \langle\hat{\theta}_{1}({\mathbf{r}},t)\hat{\theta}_{1}({\mathbf{r}+\mathbf{r'}},t)\rangle e^{-i\kvec\cdot\mathbf{r}}d\mathbf{r},
\end{equation}
where the integrand is given by \eref{eq:phase_corr2}.
In our numerical TWA implementation, we initialize the stochastic fields by adding noise to the mean field. The noise is added in the Bogoliubov mode basis and the stochastic field $\alpha({\mathbf{r}},t=0)$ is initialized as
\begin{equation}
\alpha({\mathbf{r}},t=0)=\sqrt{n_0}+\frac{1}{\sqrt{V}} \sum_{\mathbf{k},k<{k}_{max}} \Big( \beta_{{\bf{k}}} u_{{\bf{k}}} e^{i{\bf{k}}\cdot{\mathbf{r}}} + \beta^{*}_{{\bf{k}}} v_{{\bf{k}}}   e^{-i{\bf{k}}\cdot{\mathbf{r}}}\Big),
\label{eq:al_def}
\end{equation}
with $k=|\mathbf{k}|$, where $\sqrt{n_0}$ is the uniform initial wavefunction of BEC.
Here ${k}_{max}=K/2$ is the largest wavenumber for which we add noise, chosen less than the maximum $K$ allowed by our Fourier domain, to avoid aliasing.
The stochastic field $\alpha({\mathbf{r}},t)$ is then evolved according to \eref{app:GPE} with $\psi({\mathbf{r}},t) \rightarrow \alpha({\mathbf{r}},t)$.

The quantum fluctuations are captured by $\beta_{{\bf{k}}}$ which are random numbers satisfying the relation $\langle \beta_{{\bf{k}}}\rangle=0 $, $\langle \beta_{{\bf{q}}}\beta_{{\bf{k}}}\rangle=0 $ and $\langle \beta^{*}_{{\bf{q}}}\beta_{{\bf{k}}}\rangle=\delta_{{\bf{q}},{\bf{k}}} $, where $\delta_{{\bf{q}},{\bf{k}}}$ is the Kronecker delta. 

In the simulation, the wavefunction is initialized at $t=0$, and we can write 
\begin{eqnarray} \label{eq:initial_rel}
&\langle\alpha({\mathbf{r}},0)\rangle_W={\sqrt{n_0}}, \hspace{0.5cm} 
E_k = \frac{\hbar^2k^2}{2m}, \CR
&\epsilon_k=\frac{\hbar k}{\sqrt{2m}} \sqrt{E_k+\left[U_0+\frac{U_d(0)}{3}(3\cos^2\beta-1)\right]2n_0},\CR
&u_{k}=\frac{1}{2}\frac{E_k+\epsilon_k}{\sqrt{\epsilon_kE_k}}, \hspace{0.5cm}
v_{k}=\frac{1}{2}\frac{E_k-\epsilon_k}{\sqrt{\epsilon_kE_k}}.
\end{eqnarray}
Using \eref{eq:phase_corr2}-\eref{eq:initial_rel} we can analytically find the initial power spectrum as 
\begin{eqnarray}
P(k,t=0)&=\frac{1}{4n_0}[2(u_k-v_k)^2]
            =\frac{1}{2n_0}\frac{\epsilon_k}{E_k},
\end{eqnarray}
which is also used to determine initial conditions for the analytical solutions of \eref{bec_fluc_eqn}.
\vspace{1cm}
\bibliography{anisotropic}
\bibliographystyle{sebastian_v3}
\end{document}